\PassOptionsToPackage{draft}{hyperref}
\documentclass{ANStopical}
\usepackage{xstring}
\usepackage{booktabs}
\usepackage{subcaption}
\usepackage{mdframed}
\newif\ifincludeoptional
\includeoptionalfalse 
\newmdenv[linewidth=0.8pt,linecolor=black,
  leftline=true,rightline=true,topline=true,bottomline=true,everyline=false,
  innerleftmargin=4pt,innerrightmargin=4pt,
  skipabove=6pt,skipbelow=6pt]{expandednote}
\graphicspath{{figures/}}
\newcommand{\tauint}{\tau_{\mathrm{int}}}
\newcommand{\E}{\mathbb{E}}
\newcommand{\Var}{\operatorname{Var}}
\newcommand{\Cov}{\operatorname{Cov}}

\newcommand{\DomainL}{200}
\newcommand{\OpticalL}{33}
\newcommand{\ScatterRatio}{0.9}
\newcommand{\NRep}{28}
\newcommand{\NRepAbs}{40}
\newcommand{\NInitial}{1,000,000}
\newcommand{\NGen}{2000}
\newcommand{\NBurn}{200}
\newcommand{\NWindow}{1801}
\newcommand{\LagBranch}{154}
\newcommand{\LagAbs}{182}
\newcommand{\CollisionsAbs}{10,000,000}

\newcommand{\TailBranch}{0.03701}

\newcommand{\FinalMean}{1,005,272}
\newcommand{\FinalSD}{42,165}
\newcommand{\FinalRelativeSD}{4.22}
\newcommand{\PredRelativeSD}{4.47}

\newcommand{\CutoffShiftAbs}{0.02385}
\newcommand{\CutoffShiftBranch}{0.00668}

\title{An Analytic Two-Dimensional Benchmark for\\
Integrated Autocorrelation Times in Branching Monte Carlo\\
Part I: Transport Spectra and Branching Theory}
\addAuthor[jlmiao@psu.edu]{Jilang Miao}{1}
\addAuthor{Guillaume L. Giudicelli}{2}
\addAuthor{William Reed Kendrick}{3}
\addAffiliation{1}{Department of Nuclear Engineering,
Pennsylvania State University, University Park, PA 16802}
\addAffiliation{2}{Idaho National Laboratory, Idaho Falls, ID 83415}
\addAffiliation{3}{Department of Nuclear Science and Engineering,
Massachusetts Institute of Technology, Cambridge, MA 02139}
\shortTitle{Branching Monte Carlo Benchmark: Theory}
\authorHead{J. Miao et al.}

\Abstract{Predicting correlations between source generations can support uncertainty
estimation and variance reduction in Monte Carlo transport. We derive a two-dimensional
benchmark for planar isotropic exponential flights in a reflecting square.
An image construction gives the exact collision and absorption spectra.
Critical branching preserves the mean transport operator but changes mode
covariances through the offspring second moment. These covariances give
tally autocorrelations and integrated autocorrelation times, with a
large-population, finite-window approximation for normalized source tallies.
Direct Monte Carlo measurements test the predictions for absorption/rebirth
and two-offspring fission. The benchmark provides a reference for assessing
reduced correlation models, for correcting reported tally uncertainties,
and for checking the tallies of a production code for correlation bias.}

\begin{document}

\section{Introduction}

Monte Carlo eigenvalue simulations use the fission sites produced in one
generation to construct the source for the next generation. This dependence
introduces correlation between generation tallies. Treating these tallies as
independent can underestimate the variance of their average and obscure the
number of generations needed to reach a target uncertainty~\cite{Miao2016}.
Correlation prediction is therefore useful before the simulation has produced
a long tally history. Its purpose is not limited to correcting a reported
variance: the predicted behavior can also be used to evaluate source-update
schemes intended to improve statistical convergence~\cite{Miao2020}.

The effect of this correlation on the tallies of Monte Carlo eigenvalue calculations
has been an open question for decades.
A benchmark with known correlation shows for which tallies the correlation matters.
It supports corrections to reported tally uncertainties,
offers a testbed for algorithms that reduce correlation bias,
and allows the tallies of a production code to be checked for bias of unknown origin.

Previous work has investigated variance underestimation for fission-rate
distributions~\cite{Yamamoto2014} and developed multitype branching process
(MBP) predictions~\cite{Miao2018}.
Studies of spatial clustering in the power iteration~\cite{Nowak2016,Sutton2017}
are the closest prior work on the spatial structure of these correlations;
the present benchmark quantifies their consequence at the tally level.
The MBP approach accounts for source transport and offspring fluctuations,
but its discrete phase-space representation becomes costly as the number of
cells increases. Functional expansion tallies represent spatial distributions
through basis coefficients~\cite{Griesheimer2006}. Applying the same idea to
transport operators gives a natural way to investigate reduced correlation
models. To assess such models, the effect of finite representation must be
distinguished from Monte Carlo sampling error and from approximations in the
reference correlation calculation.

An earlier study examined directly tallied reduced operators for a reflective
scattering chain~\cite{Miao2026FET}. This paper develops the transport
derivation and extends the benchmark to absorption and branching. The collision, absorption, and branching
processes are related explicitly, allowing the transport and multiplication
contributions to correlation to be examined separately. A complete reflective
eigenbasis is derived by unfolding the transport kernel, rather than assumed
from a diffusion equation.
The adjective analytic refers to this spectrum:
the eigenfunctions and eigenvalues of the collision and absorption kernels
under reflective boundaries are in closed form.
The correlation references built from it involve a large-population,
finite-window approximation for the normalized branching tally,
a finite spatial cutoff of the mode sums,
and finite sampling in the direct Monte Carlo comparison.
The resulting mode covariances are then compared
with direct Monte Carlo estimates. The companion Part II uses this benchmark
to assess operators tallied directly in cell and functional bases. The present
paper establishes the reference problem and the conditions under which its
branching correlation formula applies.

\section{Transport and Correlation Theory}

\subsection{Transport and Source Processes}

Consider the square $\Omega=[-L,L]^2$ with specularly reflecting boundaries.
A neutron is emitted isotropically in the plane and travels an exponentially
distributed distance with rate $\Sigma_t$. At each collision it scatters with
probability $c_s=\Sigma_s/\Sigma_t$ and otherwise is absorbed. Scattering
resamples an isotropic direction. The cross sections are spatially constant,
with $\Sigma_t=\Sigma_s+\Sigma_a$ and $\Sigma_a=\Sigma_c+\Sigma_f$.
This is a planar transport model, not a spatial projection of a
three-dimensional angular distribution.

Let $K(\mathbf x,\mathbf y)$ be the density of the next collision position
$\mathbf y$ given an isotropic start at $\mathbf x$. 
The first argument is the departure point and the operator acts on functions:
$(Kf)(\mathbf x)=\int_\Omega K(\mathbf x,\mathbf y)f(\mathbf y)
\,d\mathbf y$. The birth-to-absorption operator is denoted by $A$.
Both operators can be shown to preserve the uniform probability density
$\pi(\mathbf x)=1/(4L^2)$. Inner products are
$\langle f,g\rangle_\pi=\int_\Omega f(\mathbf x)g(\mathbf x)
\pi(\mathbf x)\,d\mathbf x$.

Two test cases will use $A$. In the absorption/rebirth chain, one neutron
is born at each absorption site. In the branching process, an absorbed parent
produces $\xi_a$ children at that site; their subsequent histories are
independent conditional on the source bank. Capture corresponds to
$\xi_a=0$. The offspring law is independent of position and transport, with
\begin{equation}
 \E\xi_a=1,\qquad m_2=\E\xi_a^2<\infty,\qquad
 \Var(\xi_a)=m_2-1.
 \label{eq:offspring}
\end{equation}
Thus the process is critical in expectation, and no population control is applied.
The
initial bank contains $N_0$ independent uniform positions. For
$I=[-\alpha L,\alpha L]^2$, write $Z_I(g)$ for the number of source particles
in $I$ at generation $g$, $Z_W(g)$ for the total bank size, and
$X_I(g)=Z_I(g)/Z_W(g)$ when the bank is nonempty. The stationary spatial
probability of the tally region is $\pi_I=\alpha^2$.
Physically, $\Omega$ is a homogeneous planar square with mirror walls,
the source bank holds the neutrons born at fission or rebirth sites,
and a generation is one birth-to-absorption step of every particle in the bank.
The tally region $I$ is a centered square window of side $2\alpha L$,
and its source fraction $X_I(g)$ is the generation-wise source tally
whose correlation is studied.

\subsection{Reflective Transport Spectrum}

We now determine the spectrum of the collision transport operator $K$.
Earlier work derived an analytic correlation model for a one-dimensional
homogeneous problem with reflective boundaries and extended the analysis to
three dimensions using a diffusion approximation~\cite{PhdThesisJMiao,Miao2019}.
Here we extend the reflective transport-kernel analysis from one dimension to
two-dimensional planar isotropic flights, without a diffusion approximation.
We first construct the reflected kernel and then find its eigenfunctions and
eigenvalues.

In an unbounded medium the joint probability of next collision at flight length in $(s,s+ds)$ and direction
in $(\theta,\theta+d\theta)$ is
$\Sigma_t e^{-\Sigma_t s}ds\,d\theta/(2\pi)$. Dividing by the polar area
element gives the displacement density
\begin{equation}
 K_\infty(\mathbf s)=\frac{\Sigma_t e^{-\Sigma_t|\mathbf s|}}
 {2\pi|\mathbf s|}.
 \label{eq:free}
\end{equation}
Reflection folds a straight
 flight in the unbounded medium back into the square. The images of
$\mathbf y=(y_1,y_2)$ are
$\mathbf y_{j\ell}=((-1)^j y_1+2jL,(-1)^\ell y_2+2\ell L)$, so
\begin{equation}
 K(\mathbf x,\mathbf y)=\sum_{j,\ell\in\mathbb Z}
 K_\infty(\mathbf y_{j\ell}-\mathbf x).
 \label{eq:images}
\end{equation}
The image squares tile the plane except for boundaries of zero area. Therefore
integrating Eq.~\eqref{eq:images} over $\mathbf y\in\Omega$ gives unit mass.
Reciprocity of the reflected paths gives a symmetric kernel, and hence the
uniform stationary measure.

To find the spectrum of this kernel, we first construct a complete spatial
basis and then evaluate the action of $K$ on its members. Define the
one-dimensional functions
\begin{equation}
 \varphi_0(x)=1,\qquad
 \varphi_n(x)=\sqrt{2}\cos\!\left(\frac{n\pi(x+L)}{2L}\right)
 \quad(n\geq1),\qquad k_n=\frac{n\pi}{2L}.
 \label{eq:basis}
\end{equation}
The change of variable $u=(x+L)/(2L)$ maps the uniform measure $dx/(2L)$
to $du$ on $[0,1]$, where these functions are the normalized Fourier
cosine basis. Thus $e_{np}(x_1,x_2)=\varphi_n(x_1)\varphi_p(x_2)$
form a complete orthonormal basis in $L^2(\pi)$, including functions odd
about the center. The identity
$e_{np}(\mathbf y_{j\ell})=e_{np}(\mathbf y)$ unfolds the reflected
integral into a convolution over the plane:
\begin{equation}
 (Ke_{np})(\mathbf x)
 =\int_{\mathbb R^2}K_\infty(\mathbf z-\mathbf x)e_{np}(\mathbf z)
   \,d\mathbf z
 =\mu_{np}e_{np}(\mathbf x).
 \label{eq:transport-summary}
\end{equation}
Each cosine product is a sum of plane waves with the same wavenumber
$k=\sqrt{k_n^2+k_p^2}$. Radial symmetry gives them a common multiplier.
Integrating over the flight length and then the direction gives
\begin{equation}
 \mu_{np}
 =\frac{\Sigma_t}{2\pi}\int_0^{2\pi}
   \frac{d\theta}{\Sigma_t-ik\cos\theta}
 =\frac{\Sigma_t}{\sqrt{\Sigma_t^2+k^2}}.
 \label{eq:mu}
\end{equation}
In particular, $\mu_{00}=1$ and $0<\mu_{np}<1$ for nonconstant modes.
This eigensystem follows from the transport kernel without a spatial
discretization or diffusion approximation.


\subsection{Absorption and the Mean Fission Operator}

The number of flights $J$ before absorption is geometric:
$\Pr(J=j)=(1-c_s)c_s^{j-1}$ for $j\geq1$. Homogeneity makes these reaction
marks independent of the flight trajectory. Conditional on $J=j$, the
absorption position has kernel $K^j$. Consequently,
\begin{equation}
 A=(1-c_s)\sum_{j=1}^\infty c_s^{j-1}K^j
   =(1-c_s)K(\mathrm{Id}-c_sK)^{-1}.
 \label{eq:absop}
\end{equation}
Here $\mathrm{Id}$ denotes the identity operator, $\mathrm{Id}f=f$;
the notation distinguishes it from the tally region $I$.
The series converges in operator norm for $c_s<1$, because $K$ is a
contraction in $L^2(\pi)$. Since $K^j e_{np}=\mu_{np}^j e_{np}$,
applying Eq.~\eqref{eq:absop} to an eigenfunction gives
\begin{equation}
 \begin{aligned}
 Ae_{np}
 &=(1-c_s)\sum_{j=1}^\infty c_s^{j-1}\mu_{np}^j e_{np}
 =(1-c_s)\mu_{np}\sum_{j=0}^\infty(c_s\mu_{np})^j e_{np}
 =\Lambda_{np}e_{np},\\
 \Lambda_{np}
 &=\frac{(1-c_s)\mu_{np}}{1-c_s\mu_{np}}
 =\frac{\Sigma_a}{\sqrt{\Sigma_t^2+k^2}-\Sigma_s}.
 \end{aligned}
 \label{eq:lambda}
\end{equation}
Thus the collision eigenvalues $\mu_{np}$ and absorption eigenvalues
$\Lambda_{np}$ correspond to the same spatial eigenfunctions $e_{np}$.
The difference is whether a step ends at the next collision or at the next
absorption. 
The coefficients $(1-c_s)c_s^{j-1}$ in Eq.~\eqref{eq:absop} are the probabilities $\Pr(J=j)$ and sum to
one, 
so $A$ is a mixture of the normalized, $\pi$-preserving operators $K^j$, and therefore it is normalized, preserves $\pi$  and has $\Lambda_{00}=1$.


To relate this absorption operator to branching, consider one parent born at
$\mathbf x$. Let $\mathbf Y$ be its absorption position and $\xi_a$ the
number of children born there. For a test function $f$, its contribution
to the next source tally is $\xi_a f(\mathbf Y)$. Independence of the
offspring law and transport gives
\begin{equation}
 \begin{aligned}
 \E[\xi_a f(\mathbf Y)\mid\mathbf x]
 &=\E[\xi_a]\int_\Omega A(\mathbf x,\mathbf y)f(\mathbf y)\,d\mathbf y
 =(Af)(\mathbf x),\\
 \E[(\xi_a f(\mathbf Y))^2\mid\mathbf x]
 &=\E[\xi_a^2]\int_\Omega A(\mathbf x,\mathbf y)f(\mathbf y)^2\,d\mathbf y
 =m_2(Af^2)(\mathbf x).
 \end{aligned}
 \label{eq:parent-moments}
\end{equation}
The mean-contribution identity uses criticality, $\E\xi_a=1$. Hence the expected
offspring contribution is the same as for exactly one child at every
absorption, even though the realized offspring count can differ.
If $\eta_g(\mathbf x)$ denotes the expected source number density at
generation $g$, summing these mean contributions gives
\begin{equation}
 \eta_{g+1}(\mathbf y)=\int_\Omega
 \eta_g(\mathbf x)A(\mathbf x,\mathbf y)\,d\mathbf x.
 \label{eq:mean-source}
\end{equation}
Starting from $\eta_0=N_0\pi$ therefore gives $\eta_g=N_0\pi$ at every
generation because $\pi$ is invariant under $A$. This is a statement about
the ensemble mean, not a fixed bank size in each realization. The second
moment in Eq.~\eqref{eq:parent-moments} contains $m_2$, which can exceed one
in a critical branching process. Thus equal mean transport operators do not
give equal tally variances or correlations. The branching derivation below
calculates this additional contribution.

\subsection{Tally Correlations}

We first relate the correlation of a stationary tally sequence to the
variance of its average. We then calculate that correlation for the
collision and absorption/rebirth chains. These stationary-chain results
are exact. Their use for a normalized branching source requires the
finite-window approximation derived below for branching.

\subsubsection{Finite-Run Variance and Integrated Correlation}

Let $X_I(g)$ be a stationary scalar tally sequence with finite, nonzero
variance $c_0$. Define its lag covariance, correlation coefficient, and
sample average by
\begin{equation}
 c_d=\Cov[X_I(g),X_I(g+d)],\qquad
 \rho_I(d)=\frac{c_d}{c_0},\qquad
 \overline X_N=\frac{1}{N}\sum_{g=0}^{N-1}X_I(g).
 \label{eq:correlation-definition}
\end{equation}
Counting observation pairs at each lag gives the finite-run variance:
\begin{equation}
 \Var(\overline X_N)=\frac{c_0}{N}r_I(N),\qquad
 r_I(N)=1+2\sum_{d=1}^{N-1}(1-d/N)\rho_I(d).
 \label{eq:variance}
\end{equation}
Here $r_I(N)$ is the variance ratio relative to independent observations
with the same marginal variance. The factor $1-d/N$ accounts for the
number of pairs at lag $d$.

If $\sum_{d\geq1}|\rho_I(d)|<\infty$, the integrated autocorrelation
time (IACT) is
\begin{equation}
 \tauint\equiv\lim_{N\to\infty}r_I(N)
 =1+2\sum_{d=1}^\infty\rho_I(d).
 \label{eq:iact-definition}
\end{equation}


Thus $c_0\tauint$ is the asymptotic
variance coefficient of the sample average, as in standard Markov chain
Monte Carlo analysis~\cite{Geyer1992}.

The full correlation curve gives the finite-run variance through
Eq.~\eqref{eq:variance}; IACT alone gives only its large-$N$ limit.
Correlation prediction can also guide variance reduction. For example,
earlier work used a predictive model to select parameters for delayed-source
updates~\cite{Miao2020}. An approximate model may help identify which
modes or lags to target, but the modified scheme must be assessed using its
tally variance and computational cost as well as its correlations.
Developing such schemes is a motivation for this work.

\subsubsection{Indicator Tally and Spatial Coefficients}

For the region $I=[-\alpha L,\alpha L]^2$, with $0<\alpha<1$, define
\begin{equation}
 f_I(\mathbf x)=\mathbf1_I(\mathbf x)
 =\begin{cases}1,&\mathbf x\in I,\\0,&\mathbf x\notin I,\end{cases}
 \qquad h(\mathbf x)=f_I(\mathbf x)-\pi_I,\qquad \pi_I=\alpha^2.
 \label{eq:indicator}
\end{equation}
For a single particle at $\mathbf R_g$, the tally is
$X_I(g)=f_I(\mathbf R_g)$. For a nonempty source bank it is the average
of the same indicator over all particles:
\begin{equation}
 X_I(g)=\frac{1}{Z_W(g)}\sum_{p=1}^{Z_W(g)}f_I(\mathbf x_p)
 =\frac{Z_I(g)}{Z_W(g)},\qquad Z_W(g)>0.
 \label{eq:bank-average-tally}
\end{equation}
Thus the same spatial function $f_I$ defines the single-particle tally and the bank-average tally.

Write $\kappa=(n,p)$ and let $\kappa=0$ denote $(0,0)$.
Since $e_{00}=1$, expansion of $f_I$ in the orthonormal basis gives
\begin{equation}
 f_I=\pi_I+\sum_{\kappa\ne0}T_\kappa e_\kappa,\qquad
 h=\sum_{\kappa\ne0}T_\kappa e_\kappa,\qquad
 T_{np}=\langle f_I,e_{np}\rangle_\pi=T_nT_p.
 \label{eq:tally-expansion}
\end{equation}
The last factorization follows from the square tally and the uniform
product measure. Direct one-dimensional integration gives
\begin{equation}
 T_n=\frac{1}{2L}\int_{-\alpha L}^{\alpha L}\varphi_n(x)\,dx,\qquad
 T_0=\alpha,\qquad
 T_n=\frac{2\sqrt{2}}{n\pi}\cos(n\pi/2)\sin(n\pi\alpha/2)\quad(n\geq1).
 \label{eq:tally}
\end{equation}
Odd indices vanish because the tally is symmetric.
For a single uniform particle, Parseval's identity gives
\begin{equation}
 c_0=\langle h,h\rangle_\pi
 =\sum_{\kappa\ne0}T_\kappa^2=\pi_I(1-\pi_I).
 \label{eq:indicator-variance}
\end{equation}
The bank-average variance additionally depends on the correlations between
particles and is derived separately below.

\subsubsection{Pure-Scattering Collision Chain}

For the stationary collision chain, conditioning on the current position
and using the eigenbasis gives~\cite{Geyer1992}
\begin{equation}
 c_d^{\mathrm{coll}}
 =\langle h,K^dh\rangle_\pi
 =\sum_{\kappa\ne0}T_\kappa^2\mu_\kappa^d.
 \label{eq:markov-covariance}
\end{equation}

Normalizing by $c_0=\pi_I(1-\pi_I)=\sum_{\kappa\ne0}T_\kappa^2$
from Eq.~\eqref{eq:indicator-variance} gives the correlation coefficient:
\begin{equation}
 \rho_I^{\mathrm{coll}}(d)
 =\frac{c_d^{\mathrm{coll}}}{c_0}
 =\frac{\sum_{\kappa\ne0}T_\kappa^2\mu_\kappa^d}
        {\pi_I(1-\pi_I)}
 =\frac{\sum_{\kappa\ne0}T_\kappa^2\mu_\kappa^d}
        {\sum_{\kappa\ne0}T_\kappa^2}.
 \label{eq:acf}
\end{equation}

Summing the geometric lag series for each nonconstant mode gives
\begin{equation}
 \tauint^{\mathrm{coll}}
 =1+2\frac{\sum_{\kappa\ne0}T_\kappa^2
                      \mu_\kappa/(1-\mu_\kappa)}
                  {\sum_{\kappa\ne0}T_\kappa^2}.
 \label{eq:tau}
\end{equation}
Absolute summability holds here because every nonconstant eigenvalue is
bounded by $\mu_{10}<1$ and $\sum_{\kappa\ne0}T_\kappa^2=c_0<\infty$.
At finite $N$, the same correlation curve is inserted into
Eq.~\eqref{eq:variance}, retaining its lag-dependent weights.

\subsubsection{Absorption with One-Particle Rebirth}

If every absorption is followed by exactly one birth at the same position,
$\xi_a=1$ and the recorded positions form a Markov chain with transition
operator $A$. Its stationary distribution is still $\pi$. Replacing $K$
by $A$ in Eq.~\eqref{eq:markov-covariance} and using their common
eigenfunctions gives
\begin{align}
 c_d^{\mathrm{abs}}
 &=\langle h,A^dh\rangle_\pi
 =\sum_{\kappa\ne0}T_\kappa^2\Lambda_\kappa^d,\qquad
 \rho_I^{\mathrm{abs}}(d)
 =\frac{\sum_{\kappa\ne0}T_\kappa^2\Lambda_\kappa^d}
        {\sum_{\kappa\ne0}T_\kappa^2},\label{eq:acf-abs}\\
 \tauint^{\mathrm{abs}}
 &=1+2\frac{\sum_{\kappa\ne0}T_\kappa^2
                      \Lambda_\kappa/(1-\Lambda_\kappa)}
                  {\sum_{\kappa\ne0}T_\kappa^2}.
 \label{eq:tau-abs}
\end{align}
Only the eigenvalues change from the collision-chain formulas. The tally
coefficients and single-particle variance remain the same.

Since $\Lambda_\kappa/(1-\Lambda_\kappa)
=(1-c_s)\mu_\kappa/(1-\mu_\kappa)$ for every nonconstant mode,
the two event clocks are related by Eq.~\eqref{eq:clock} below.

\begin{equation}
 \tauint^{\mathrm{abs}}-1=(1-c_s)(\tauint^{\mathrm{coll}}-1).
 \label{eq:clock}
\end{equation}
This relation holds for the integrated correlations of the same tally observed at collisions or at absorptions; the lag-by-lag $\rho_I(d)$ transforms differently, and branching requires the separate treatment below.

For $N_0$ independent stationary particles, averaging divides every lag
covariance by $N_0$, leaving the normalized ACF and IACT unchanged.
This comparison does not apply to a branching bank, whose particles can
share ancestors.

\subsubsection{Critical Branching and Mode Covariance}

In a branching bank, correlations between descendants prevent us from
obtaining the tally covariance by simply dividing the single-particle
result by the population size. We instead follow the spatial fluctuations
of the whole bank. The eigenfunctions are useful for this purpose because
$Ae_\kappa=\Lambda_\kappa e_\kappa$: mean propagation acts separately
on each mode, while the offspring moments determine its fluctuations.
The indicator coefficients $T_\kappa$ then relate these mode fluctuations
to the tally of interest.

Let $\mathcal F_g$ denote the information in the source-bank history through
generation $g$. For a nonconstant mode $\kappa$, define the bank's mode
amplitude as the sum of its eigenfunction values over all current particles:
\begin{equation}
 Z_g^{(\kappa)}=\sum_{p=1}^{Z_W(g)}e_\kappa(\mathbf x_p).
 \label{eq:mode-amplitude}
\end{equation}
This is an unnormalized spatial tally of the bank, not the amplitude of
one particle. Its expectation is zero for the uniform mean source because
$\langle1,e_\kappa\rangle_\pi=0$.
Summing the expansion of $f_I$ in Eq.~\eqref{eq:tally-expansion} over the bank gives
\begin{equation}
 Z_I(g)-\pi_I Z_W(g)
 =\sum_{\kappa\ne0}T_\kappa Z_g^{(\kappa)}.
 \label{eq:bank-tally-modes}
\end{equation}
Thus the tally covariance follows from the mode variances (Eq.~\eqref{eq:varrec}) and lag
covariances (Eq.~\eqref{eq:mode-lag-covariance}). The random population denominator is treated separately below.

One parent at $\mathbf x_p$ is absorbed at $\mathbf y_p$ and contributes
$\xi_p e_\kappa(\mathbf y_p)$ to the next bank. Conditional independence
between parents and Eq.~\eqref{eq:offspring} give
\begin{align}
 \E[Z_{g+1}^{(\kappa)}\mid\mathcal F_g]
   &=\Lambda_\kappa Z_g^{(\kappa)},\label{eq:drift}\\
 \Var[Z_{g+1}^{(\kappa)}\mid\mathcal F_g]
   &=\sum_p\left[m_2(Ae_\kappa^2)(\mathbf x_p)
                -\Lambda_\kappa^2e_\kappa(\mathbf x_p)^2\right].
 \label{eq:condvar}
\end{align}
The raw moment $m_2$ appears because all children of one parent are born
at the same site: their squared contribution is
$\xi_p^2 e_\kappa(\mathbf y_p)^2$. Conditional independence is required
between parents' subsequent histories, not between particles in the
current bank. Write $v_{g,\kappa}=\Var[Z_g^{(\kappa)}]$.

The uniform mean source gives
$\E[\sum_p f(\mathbf x_p)]=N_0\langle1,f\rangle_\pi$.
Since $\pi$ is invariant and $e_\kappa$ has unit norm,
$\langle1,A(e_\kappa^2)\rangle_\pi
=\langle1,e_\kappa^2\rangle_\pi=1$.
Averaging Eq.~\eqref{eq:condvar} therefore gives
\begin{equation}
 Q_\kappa\equiv\E[\Var(Z_{g+1}^{(\kappa)}\mid\mathcal F_g)]
 =N_0(m_2-\Lambda_\kappa^2).
 \label{eq:mode-variance-input}
\end{equation}


The law of total variance then gives the recurrence
\begin{equation}
 v_{g+1,\kappa}=\Lambda_\kappa^2v_{g,\kappa}+Q_\kappa
 =\Lambda_\kappa^2v_{g,\kappa}+N_0(m_2-\Lambda_\kappa^2).
 \label{eq:varrec}
\end{equation}

Independent uniform initial
positions give $v_{0,\kappa}=N_0$, and $0<\Lambda_\kappa<1$ gives
the limiting nonconstant-mode variance
\begin{equation}
 v_{\infty,\kappa}
 =N_0\frac{m_2-\Lambda_\kappa^2}{1-\Lambda_\kappa^2}.
 \label{eq:equilibrated-mode-variance}
\end{equation}

We also need the covariances between different modes to reconstruct the
tally covariance. Define
$C_{g,\kappa\ell}=\Cov(Z_g^{(\kappa)},Z_g^{(\ell)})$.
The same conditional-moment calculation, with $e_\kappa e_\ell$ in
place of $e_\kappa^2$, and the law of total covariance give
\begin{equation}
 C_{g+1,\kappa\ell}
 =\Lambda_\kappa\Lambda_\ell C_{g,\kappa\ell}
   +N_0(m_2-\Lambda_\kappa\Lambda_\ell)\delta_{\kappa\ell}.
 \label{eq:cross-mode-recurrence}
\end{equation}
Here $\langle e_\kappa,e_\ell\rangle_\pi=\delta_{\kappa\ell}$.
For distinct modes, the added term vanishes by orthogonality, and the
independent uniform initial positions give $C_{0,\kappa\ell}=0$.
Thus distinct modes remain uncorrelated at second order; they need not
be independent or Gaussian.

Finally, repeated conditioning in Eq.~\eqref{eq:drift} gives the lag
covariances:
\begin{equation}
 \begin{aligned}
 \E[Z_{g+d}^{(\ell)}\mid\mathcal F_g]
 &=\Lambda_\ell^d Z_g^{(\ell)},\\
 \Cov(Z_g^{(\kappa)},Z_{g+d}^{(\ell)})
 &=\Cov\!\left(Z_g^{(\kappa)},
          \E[Z_{g+d}^{(\ell)}\mid\mathcal F_g]\right)
 =\Lambda_\ell^d C_{g,\kappa\ell}.
 \end{aligned}
 \label{eq:mode-lag-covariance}
\end{equation}
For equal modes this is $\Lambda_\kappa^d v_{g,\kappa}$; for
distinct modes it is zero. These results supply the mode covariances
needed in Eq.~\eqref{eq:bank-tally-modes} without introducing an
independent-noise assumption.

The constant amplitude is $Z_W$. Subtracting the uniform tally fraction
removes it exactly from the numerator:
\begin{equation}
 X_I(g)-\pi_I=\frac{Z_I(g)-\pi_I Z_W(g)}{Z_W(g)}
 =\frac{\sum_{\kappa\ne0}T_\kappa Z_g^{(\kappa)}}{Z_W(g)}
 \simeq\frac{1}{N_0}\sum_{\kappa\ne0}T_\kappa Z_g^{(\kappa)}.
 \label{eq:normalized}
\end{equation}
The final step is an approximation. After nonconstant mode variances have
equilibrated, substitution of Eq.~\eqref{eq:equilibrated-mode-variance} into the covariance
of Eq.~\eqref{eq:normalized} gives
\begin{equation}
 W_\kappa^{\mathrm{br}}=
 \frac{m_2-\Lambda_\kappa^2}{1-\Lambda_\kappa^2}T_\kappa^2,\qquad
 \rho_I^{\mathrm{br}}(d)\simeq
 \frac{\sum_{\kappa\ne0}W_\kappa^{\mathrm{br}}\Lambda_\kappa^d}
      {\sum_{\kappa\ne0}W_\kappa^{\mathrm{br}}}.
 \label{eq:branchweights}
\end{equation}
Summing the lag series of this approximated, equilibrated covariance gives
\begin{equation}
 \tauint^{\mathrm{br}}\simeq
 1+2\frac{\sum_{\kappa\ne0}W_\kappa^{\mathrm{br}}
                     \Lambda_\kappa/(1-\Lambda_\kappa)}
             {\sum_{\kappa\ne0}W_\kappa^{\mathrm{br}}}.
 \label{eq:tau-branch}
\end{equation}

Setting $m_2=1$ gives $W_\kappa^{\mathrm{br}}=T_\kappa^2$ and recovers
the absorption/rebirth result. For $m_2>1$, the additional
factor $(m_2-1)/(1-\Lambda_\kappa^2)$ amplifies slow modes. The mean
transport spectrum alone cannot describe this effect.

An uncontrolled critical branching population has no stationary
finite-population law. In particular,
$\Var[Z_W(g)]=gN_0(m_2-1)$, so the relative population scale is
$\sqrt{g(m_2-1)/N_0}$. The approximation requires a window long enough for
the contributing nonconstant modes to equilibrate but short enough that
this population scale remains small.  Population control would change the offspring
dependence and requires a separate analysis, as in the normalized-tally
approximations used in MBP prediction~\cite{Miao2018}.

\section{Direct Monte Carlo Validation}

The simulations use $L=\DomainL\,\mathrm{cm}$,
$\Sigma_tL=\OpticalL$, $c_s=\ScatterRatio$, and $\alpha=1/2$.
The optical size and tally region follow the earlier scattering
benchmark~\cite{Miao2026FET}, where the optical size was chosen to represent
slow mixing comparable to the dominant mode of a large thermal reactor.
Both source processes use the same transport parameters so that their
correlation differences arise from branching.

Absorption/rebirth is measured in \NRepAbs{} independent batches, each
containing \CollisionsAbs{} collisions; the indicator sequence retains
absorption sites only. The branching calculation uses \NRep{} independent
replicas with $N_0=\NInitial$, \NGen{} generations, and offspring
$\xi_a\in\{0,2\}$ with equal probabilities. This offspring law gives
$\E\xi_a=1$ and $m_2=2$, allowing nonzero offspring variance while preserving
criticality. Source tallies from generation
\NBurn{} onward give \NWindow{} observations per replica. Operator
estimation, addressed in Part II, is not used in these direct measurements.
Independent repetitions supply the sampling uncertainties reported below;
the population size and observation window are also checked against the
population-fluctuation scale required by Eq.~\eqref{eq:normalized}.

For a replica sequence $x_0,\ldots,x_{n-1}$, the covariance estimate is
\begin{equation}
 \widehat c_d=\frac{1}{n-d}\sum_{g=0}^{n-d-1}
 (x_g-\bar x)(x_{g+d}-\bar x),\qquad
 \widehat\rho_d=\widehat c_d/\widehat c_0.
 \label{eq:directest}
\end{equation}
Each replica is centered on its own mean, not the ensemble mean. The reported
ACF and IACT are averages of replica-normalized estimates. Uncertainties are
sample standard deviations across independent replicas divided by the square
root of their number.

\begin{figure}[htbp]
 \centering
 \begin{subfigure}[t]{0.49\textwidth}
 \centering
 \includegraphics[width=\linewidth]{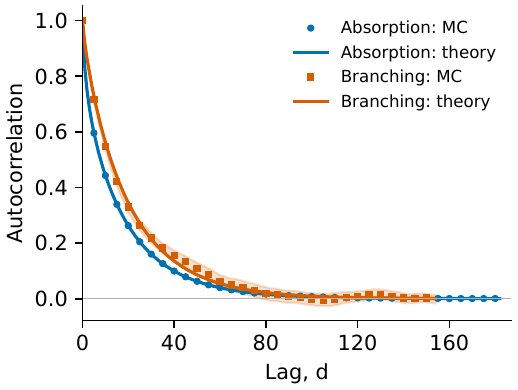}
 \caption{Autocorrelation.}
 \label{fig:direct-acf}
 \end{subfigure}\hfill
 \begin{subfigure}[t]{0.49\textwidth}
 \centering
 \includegraphics[width=\linewidth]{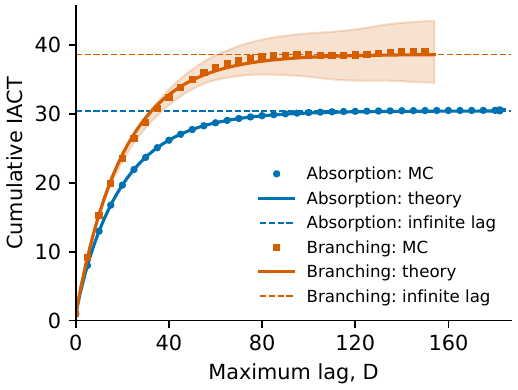}
 \caption{Cumulative IACT.}
 \label{fig:direct-iact}
 \end{subfigure}
 \caption{Direct Monte Carlo measurements and spectral predictions for
 absorption/rebirth and critical branching. Shading and the absorption IACT
 endpoint error bar give one standard error of the mean. In (b), solid lines
 sum the predicted ACF through lag $D$; dashed lines sum it to infinite lag.}
 \label{fig:direct}
\end{figure}

Figure~\ref{fig:direct-acf} compares both ACFs with their spectral predictions.
Figure~\ref{fig:direct-iact} shows their cumulative integrals,
$\widehat\tau_{\mathrm{int}}(D)=1+2\sum_{d=1}^{D}\widehat\rho_d$,
over the saved lag ranges, together with the corresponding spectral sums
and their infinite-lag limits. The branching IACT band is calculated by
summing within each replica before estimating the standard error, preserving
the dependence between lag estimates. For absorption, only the endpoint
IACT uncertainty was saved; no intermediate-lag band is inferred from the
individual ACF error bars. Finite-window centering and omitted lag tails
can introduce bias not included in these sampling uncertainties.


Branching correlations persist more strongly than those of a single
absorption/rebirth chain, despite their common mean operator.
The cumulative curves show how the integrated estimates approach the
spectral predictions as more lags are included. 


All branching trajectories remain nonempty. The final bank has ensemble
mean \FinalMean{} and sample standard deviation \FinalSD{}, or
\FinalRelativeSD\% of $N_0$. The branching population formula predicts
\PredRelativeSD\%. This comparison supports the finite-window population
scale used in Eq.~\eqref{eq:normalized}; it does not make the denominator
replacement exact. Direct branching IACT has substantially larger sampling
uncertainty than the absorption result because each normalized source
trajectory contains relatively few generations.

The spectral references retain even-sector indices $0\leq m,q\leq500$,
or full-basis indices $n=2m$, $p=2q$, in both the numerator and denominator.
Increasing the spatial cutoff to 2000 changes the infinite-lag IACT by
\CutoffShiftAbs{} for absorption and \CutoffShiftBranch{} for branching,
less than the direct Monte Carlo standard errors. These are finite-cutoff
numerical references.


\section{Conclusions}

The reflective planar transport kernel gives exact collision and absorption
spectra. Critical branching preserves the mean operator but changes mode
covariances through the raw offspring second moment. Conditional moments
determine these covariances without assuming independent particles within
the source bank.

Direct Monte Carlo measurements are consistent with the predicted integrated
correlations at their sampling precision. Normalized branching tallies
require a large-population, finite-window approximation, and the evaluated
references retain a finite spatial cutoff. These references support tests
of reduced correlation models for uncertainty prediction and source-update
design. They also show for which tallies correlation matters
and allow the tallies of a production code to be checked for bias of unknown origin.
Variance-reduction schemes and extensions to heterogeneous systems
with population control remain future work.

\bibliographystyle{ANStopical}
\bibliography{references}

@article{Geyer1992,
  author = {Charles J. Geyer},
  title = {{Practical Markov Chain Monte Carlo}},
  journal = {Statistical Science},
  volume = {7},
  number = {4},
  pages = {473--483},
  year = {1992},
  doi = {10.1214/ss/1177011137}
}

@article{Griesheimer2006,
  author = {David P. Griesheimer and William R. Martin and James Paul Holloway},
  title = {{Convergence Properties of Monte Carlo Functional Expansion Tallies}},
  journal = {Journal of Computational Physics},
  volume = {211},
  number = {1},
  pages = {129--153},
  year = {2006},
  doi = {10.1016/j.jcp.2005.05.023}
}

@article{Yamamoto2014,
  author = {Akio Yamamoto and Kotaro Sakata and Tomohiro Endo},
  title = {{Prediction on Underestimation of Variance for Fission Rate
    Distribution in Monte-Carlo Calculation}},
  journal = {Transactions of the American Nuclear Society},
  volume = {110},
  number = {1},
  pages = {515--518},
  year = {2014},
  url = {https://www.ans.org/pubs/transactions/article-35861/}
}

@article{Miao2016,
  author = {Jilang Miao and Benoit Forget and Kord Smith},
  title = {{Analysis of Correlations and Their Impact on Convergence Rates
    in Monte Carlo Eigenvalue Simulations}},
  journal = {Annals of Nuclear Energy},
  volume = {92},
  pages = {81--95},
  year = {2016},
  doi = {10.1016/j.anucene.2016.01.037}
}

@article{Miao2018,
  author = {Jilang Miao and Benoit Forget and Kord Smith},
  title = {{Predicting Correlation Coefficients for Monte Carlo Eigenvalue
    Simulations with Multitype Branching Process}},
  journal = {Annals of Nuclear Energy},
  volume = {112},
  pages = {307--321},
  year = {2018},
  doi = {10.1016/j.anucene.2017.10.014}
}

@phdthesis{PhdThesisJMiao,
  author = {Jilang Miao},
  title = {{Correlations in Monte Carlo Eigenvalue Simulations: Uncertainty
    Quantification, Prediction and Reduction}},
  school = {Massachusetts Institute of Technology},
  type = {{PhD} thesis},
  year = {2018},
  url = {https://hdl.handle.net/1721.1/123375}
}

@article{Miao2019,
  author = {Jilang Miao and Benoit Forget and Kord Smith},
  title = {{Correlation Diagnosis Method for Heterogeneous Monte Carlo
    Eigenvalue Simulations Based on a Diffusion Approximation}},
  journal = {Annals of Nuclear Energy},
  volume = {130},
  pages = {301--318},
  year = {2019},
  doi = {10.1016/j.anucene.2019.02.048}
}

@article{Miao2020,
  author = {Jilang Miao and Benoit Forget and Kord Smith},
  title = {{Improving Variance Convergence Rate in Monte Carlo Eigenvalue
    Simulations via Delayed Neutrons}},
  journal = {Annals of Nuclear Energy},
  volume = {142},
  pages = {107376},
  year = {2020},
  doi = {10.1016/j.anucene.2020.107376}
}

@article{Nowak2016,
  author = {Michel Nowak and Jilang Miao and Eric Dumonteil and Benoit Forget
    and Anthony Onillon and Kord S. Smith and Andrea Zoia},
  title = {{Monte Carlo Power Iteration: Entropy and Spatial Correlations}},
  journal = {Annals of Nuclear Energy},
  volume = {94},
  pages = {856--868},
  year = {2016},
  doi = {10.1016/j.anucene.2016.05.002}
}

@article{Sutton2017,
  author = {Thomas M. Sutton and Anudha Mittal},
  title = {{Neutron Clustering in Monte Carlo Iterated-Source Calculations}},
  journal = {Nuclear Engineering and Technology},
  volume = {49},
  number = {6},
  pages = {1211--1218},
  year = {2017},
  doi = {10.1016/j.net.2017.07.008}
}

@misc{Miao2026FET,
  author = {Jilang Miao and Guillaume L. Giudicelli and William Reed Kendrick},
  title = {{Functional Expansion Tallies of Matrix Operators for Prediction for
    Integrated Autocorrelation Time in Batch Monte Carlo: an Analytic 2D
    Scattering Chain Benchmark}},
  year = {2026},
  howpublished = {{ANS} Winter Meeting},
  note = {Accepted for publication}
}
\end{document}